\documentclass[aip,jcp,graphicx,reprint]{revtex4-1}
\usepackage{graphicx} 
\usepackage{amsmath}
\begin{document}

\title{Nudged-Elastic Band Calculations of Polymorph Transitions and Solid-State Reactions in Molecular Crystals} 

\author{Natalia Goncharova}
\author{Johannes Hoja}
\email[]{johannes.hoja@uni-graz.at}
\affiliation{Department of Chemistry, University of Graz, Heinrichstraße 28/IV, 8010 Graz, Austria.}

\date{14 June 2025}

\begin{abstract}
The modeling of solid-state transformations, such as polymorphic transitions and chemical reactions in molecular crystals, is vital for many applications including drug design or the development of new synthesis methods. However, a description via nudged-elastic band (NEB) calculations faces several crucial challenges.  First, the automatic initial pathway generation based on a linear interpolation often fails for periodic systems, leading to unrealistic geometries and atomic collisions. Second, the necessary system sizes are typically beyond the scope of density functional theory (DFT) calculations in terms of computational cost, but the associated accuracy is vitally needed. To address these issues, we introduce a hybrid interpolation method that combines linear interpolation for cell parameters with spherical linear interpolation (SLERP) for molecular structures or intramolecular fragments, ensuring smooth and realistic transitions. Moreover, we train and benchmark machine-learned force fields (MLFFs) based on the SO3krates equivariant neural network architecture to accelerate NEB calculations while retaining near-DFT accuracy. We apply our approach to two polymorph transitions and a solid-state Diels-Alder reaction and show that our new interpolation method reliably produces viable initial pathways. The MLFFs are trained on PBE+MBD reference data and reproduce DFT lattice energies with a mean absolute error of 0.4 kJ\,mol$^{-1}$ along the minimum-energy paths. These results highlight the potential of combining advanced interpolation techniques with MLFFs to enable automated, accurate, and efficient exploration of solid-state transformations in molecular crystals.
\end{abstract}

\pacs{}

\maketitle 

\section{Introduction}
Understanding solid-state transformations, such as transitions between molecular crystal polymorphs or solid-state chemical reactions at a molecular level is vital for pharmaceuticals, chemistry, and materials science\cite{Kaupp2002,Datta2004,Anthony2006,Saha2022}.
Polymorphs are different crystal-packing arrangements of the same molecule, and although the chemical composition of a molecular crystal does not change when one polymorph transforms into another, this can often be associated with quite drastic changes in several properties.

An important issue for the pharmaceutical industry is that polymorph transitions can affect the efficacy of drugs\cite{Zhou2018}, exemplified by the HIV medication ritonavir\cite{Bauer2001,Bucar2015} that became unusable after an unexpected formation of a less soluble polymorph. 
Therefore, crystal structure prediction\cite{Price2014} (CSP), which yields a polymorphic energy landscape of a given molecular crystal, has become an invaluable tool for gaining knowledge about polymorphism and the accuracy of computational approaches is regularly assessed by blind tests organized by the Cambridge Crystallographic Data Centre\cite{Reilly2016,Hunnisett2024,Hunnisett2024b}. 
The resulting thermodynamical polymorph stability rankings often include many more structures than are found experimentally since kinetic factors are not taken into account\cite{Price2013,Price2022,Beran2023}. Therefore, understanding the kinetics of polymorph transitions is key to solving this overprediction problem.
Moreover, studying chemical reactions in the solid state is becoming ever more important since these reactions, as opposed to synthesis in solution, can open up novel reaction pathways and selectivities\cite{Galwey2014}. 
However, characterizing these transformations is challenging due to the complex interplay between molecular reorganization and the constraints imposed by the crystal lattice. 

One key method for modeling solid-state transformations in molecular crystals is molecular dynamics (MD) but the relevant transitions are rarely observed in unbiased trajectories, so enhanced-sampling schemes are usually indispensable.
Several metadynamics studies\cite{Salvalaglio2014,Gimondi2017,Francia2020,Francia2021} focused on polymorph transitions of molecular crystals utilizing traditional force fields and could reduce the number of structures in CSP landscapes to a manageable set of distinct  finite-temperature basins that better reflect experimental reality.
Another promising methodology for reducing the overprediction of structures in molecular CSP is the threshold algorithm\cite{Yang2022,Butler2023} based on Monte Carlo simulations, which provides a mapping of connected structures into basins and also yields approximate inter-polymorph barriers.
However, most traditional force fields are not designed to handle the breaking of chemical bonds, which needs to occur when modeling chemical reactions, and their accuracy is often not sufficient when we are interested in the actual energy barriers (or reaction rates) of a transformation. 

The primary computational approach to investigate the barriers of reactions or transformations on a first-principles level involves identifying the minimum-energy pathway (or reaction coordinate) that connects the initial and final states of a transformation. This is typically achieved through the use of the nudged elastic band (NEB) method or similar approaches\cite{1998,Henkelman2000,Sheppard2012,Mader2018,Shin2019,Han2020,Hossain2020,Podewitz2021}, which efficiently samples the minimum-energy path by connecting a series of images or configurations along a reaction path. NEB is particularly effective for determining transition states and energy barriers, as it maintains a series of replicas of the system that are connected by springs, ensuring smooth transitions along the reaction coordinate. Therefore, this constitutes the main methodology for studying chemical reactions in vacuum or within implicit solvent models.

However, NEB calculations for solid-state systems are often hindered by the complexity of generating realistic initial reaction pathways, particularly for periodic molecular crystal structures. Linear interpolation, the simplest method for generating these paths, often fails to capture the intricate changes in both molecular orientation and unit cell parameters that occur during solid-state transformations, frequently leading to atomic collisions or nonphysical geometries. Although in isolated systems problems with the interpolation can be corrected by hand, such manual corrections are extremely cumbersome for complex periodic structures. 

The main workhorse for the modeling of molecular crystals on a first-principles level\cite{OterodelaRoza2012,Reilly2013,Marom2013,Reilly2015,Brandenburg2016,Shtukenberg2017,Hoja2019,Dolgonos2019,Jana2021,Price2023,Hoja2023} is density functional theory (DFT), since it provides a good balance between accuracy and computational cost. While first-principles MD simulations can be performed for small molecules in vacuum, modeling the kinetics of polymorph transitions and solid-state reactions with either MD or NEB for realistic systems is typically out of reach in terms of available computation times. 
In recent years, great advances have been made in the field of machine-learned force fields\cite{Chmiela2017,Schtt2017,Chmiela2018,Schutt21a,Poltavsky2021,Unke2021,Sauceda2022,Batzner2022,Batatia2022,frank2022so3krates,frank2024euclidean,Kovcs2025} (MLFFs), especially in the development of neural networks, and these methods are increasingly utilized for molecular crystals\cite{Raimbault2019,Krynski2021,Lewis2023,Zugec2024, Mladineo2024,Zhao2025,Nickerson2025}. With an appropriate set of accurate reference data --- typically high-level DFT calculations --- MLFFs can be trained so that they are almost as efficient as traditional force fields but also much more accurate for specific applications. 

In this paper, we illustrate how energy barriers of molecular crystal transformations can be obtained with near-DFT accuracy relying solely on NEB calculations. First, we present a novel interpolation method to generate high-quality initial reaction paths for NEB calculations of molecular crystals. The proposed method is designed to handle both rigid and flexible molecular systems, automatically avoiding atomic collisions and ensuring smooth transitions between the initial and final configurations. Then, we also train and benchmark the applicability and accuracy of state-of-the-art MLFFs in the context of periodic NEB calculations. For that we generate tailor-made DFT reference data and utilize the equivariant message-passing neural network architecture SO3krates\cite{frank2022so3krates,frank2024euclidean}.   

This methodology is then tested on the polymorph transitions of the $\alpha$ (low-temperature) polymorph of 1,4-diiodobenzene\cite{Alcob1994} into the $\beta$ (high-temperature) polymorph as well as that of theophylline polymorph I~\cite{Khamar2011} into polymorph II.\cite{Fucke2012} In addition, we have also modeled the solid-state reaction of a cocrystal composed of bis(N-allylimino)-1,4-dithiin and 9-bromoanthracene into the Diels-Alder product\cite{Khorasani2015}. We show that our new interpolation method automatically produces viable initial reaction paths and that the obtained MLFFs are providing very accurate results compared to the much more expensive DFT reference data for energy barriers of polymorph transitions as well as solid-state reactions.

\section{Computational Methods}

\begin{figure*}[!]
\includegraphics[width=0.75\textwidth]{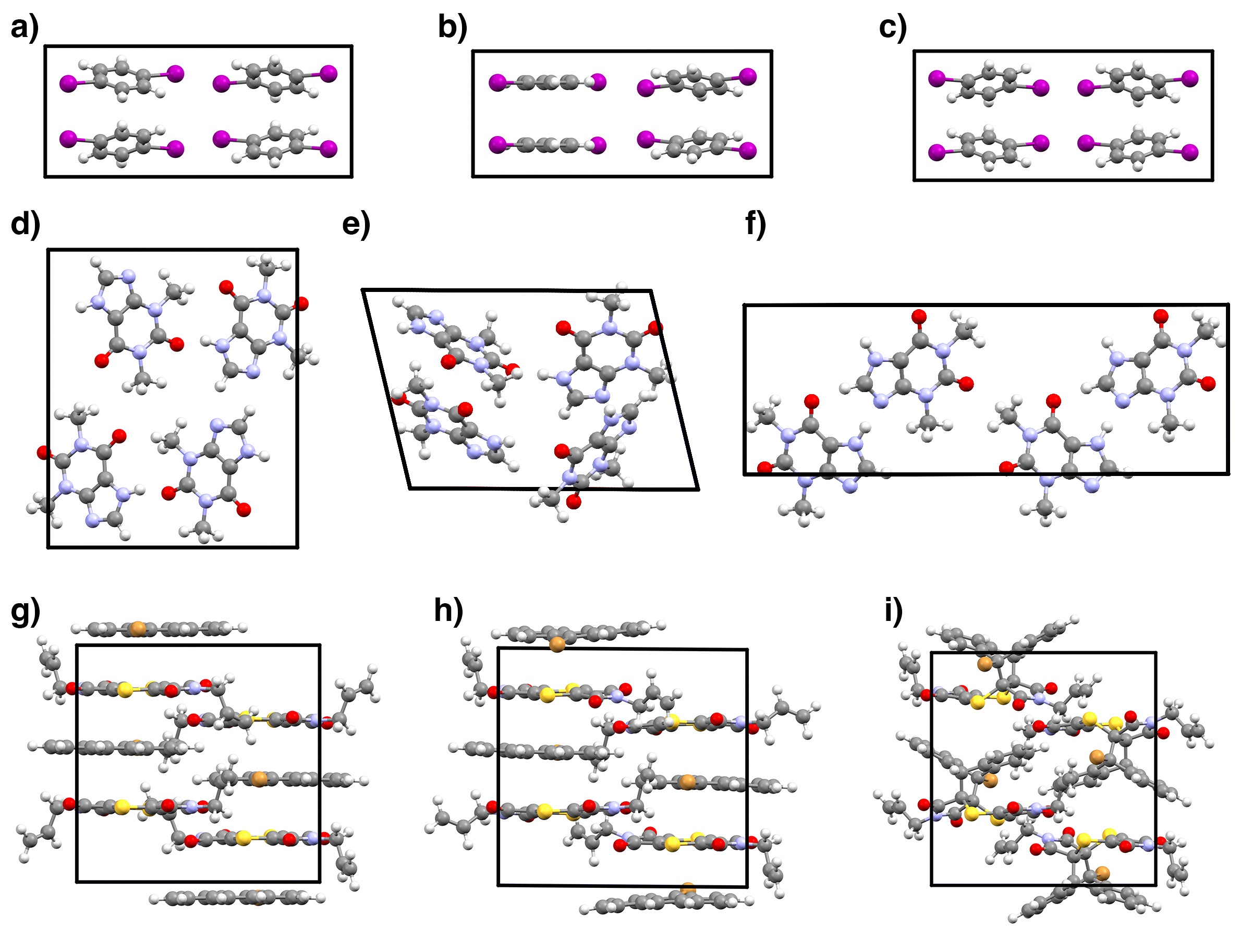}
\caption{\label{fig1} Unit cells of 1,4-diiodobenzene polymorph $\alpha$ (a), polymorph $\beta$ (c), the obtained transition state between a and c (b), theophylline polymorph I (d), polymorph II (f), the obtained transition state between d and f (e), co-crystal of bis(N-allylimino)-1,4-dithiin and 9-bromoanthracene (g), the resulting Diels-Alder product (i), and the obtained transition state between g and i (h).}
\end{figure*}

For the three considered transitions, we have utilized the following experimentally determined crystal structures as starting points, with the Cambridge structural database code listed in parenthesis:
1,4-diiodobenzene polymorph $\alpha$ (ZZZPRO05), 1,4-diiodobenzene polymorph $\beta$ (ZZZPRO07), theophylline polymorph I (BAPLOT04), 
theophylline polymorph II (BAPLOT06), co-crystal of bis(N-allylimino)-1,4-dithiin and 9-bromoanthracene (FUSCIH), and  the respective Diels-Alder product (FURROB). All these structures are visualized in Fig. \ref{fig1}.

First, all structures were fully optimized with PBE+MBD-NL\cite{Perdew1996,Tkatchenko2012, Ambrosetti2014, Hermann2020} utilizing the 2020 light species default settings for basis functions and integration grids within FHI-aims\cite{Blum2009,Knuth2015,Ren2012,Yu2018,Havu2009,Ihrig2015} (version 231212). The k-point grids for the DFT part and the MBD part were chosen so that the number of k-points $n \times l \geq 18$ \AA{} or 30 \AA{}, respectively, where $l$ is the cell length in the respective direction. We have applied the following convergence criteria: $10^{-6}$ eV for the total energy, 10$^{-5}$ electrons per \AA{}$^{3}$ for the charge density, 10$^{-4}$ eV\,\AA{}$^{-1}$ for forces, 10$^{-3}$ eV for the sum of the eigenvalues, and 0.005 eV\,\AA{}$^{-1}$ during geometry optimizations for the maximal force component. All calculations were carried out via the Atomic Simulation Environment ASE\cite{ase-paper}. 

Since a simple linear interpolation does not allow the generation of reasonable starting structures for a reaction path sampling via nudged elastic band (NEB) calculations for all investigated systems, we have developed the following interpolation methodology.
Our approach to NEB pathway generation introduces a hybrid interpolation method, which uses a linear interpolation for the unit cell and leverages the strengths of the spherical linear interpolation algorithm\cite{Shoemake1985} (SLERP) for molecular structures 
\begin{equation}
    \mathrm{Slerp}(q_1, q_2, u) = \frac{\sin{(1-u)} \Theta}{\sin{\Theta}} q_1 + \frac{\sin{u \Theta}}{\sin{\Theta}}q_2,
\end{equation} 
with $q_1$ and $q_2$ being the quaternions representing the initial and final structures, and $u$ being the interpolation parameter between 0 and 1. SLERP was, for example, already utilized for protein calculations\cite{Nguyen2017,Nguyen2018}.

\begin{figure}[!]
\includegraphics[width=\columnwidth]{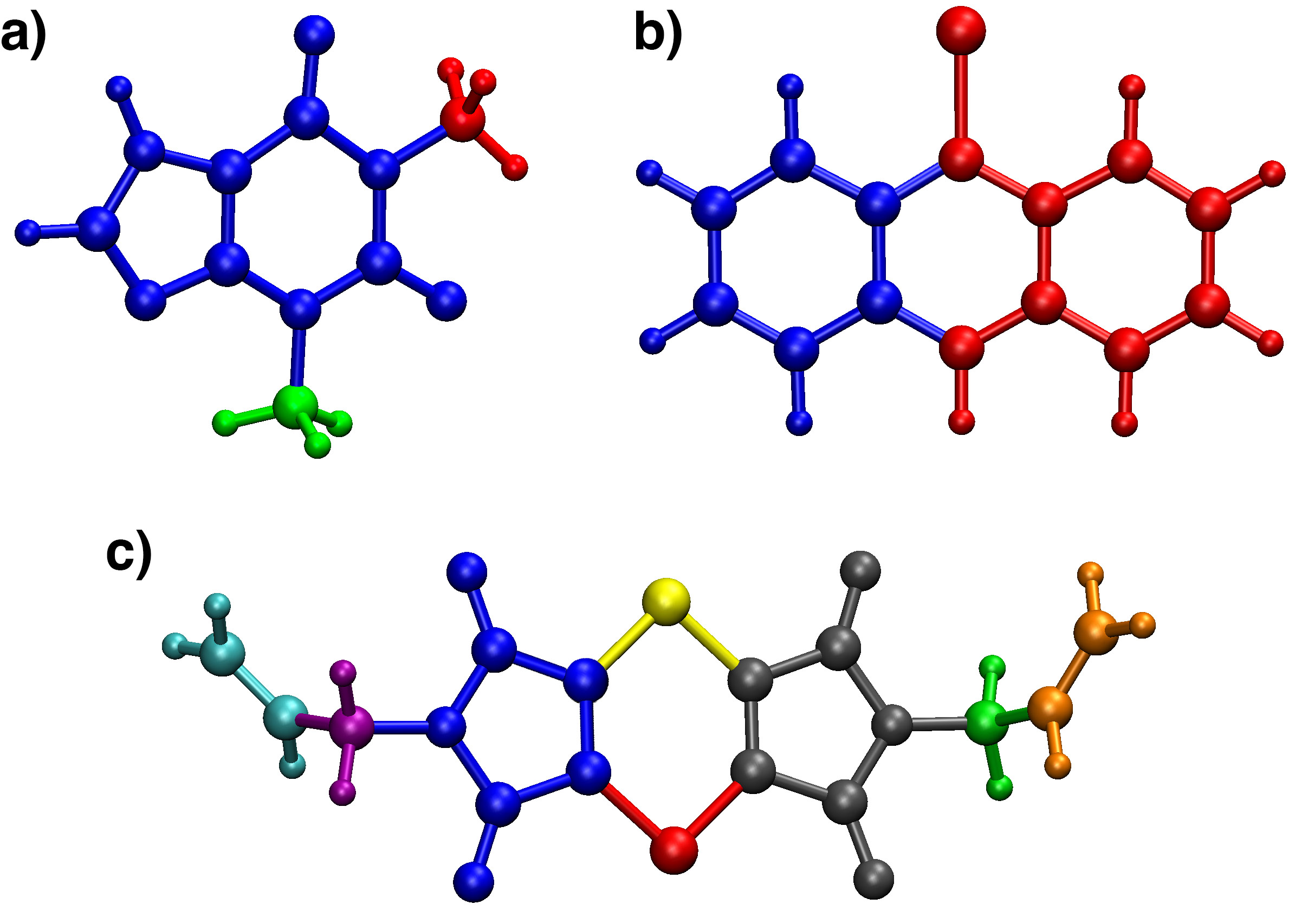}
\caption{\label{fig2}Visualization of the fragment definition (shown in different colors) used for the initial reaction-path interpolation for theophylline (a), 9-bromoanthracene (b), and bis(N-allylimino)-1,4-dithiin (c).}
\end{figure}

To properly capture the changes of molecular structures during the transformation, it is necessary to specify bonds that are likely to rotate during the reaction. 
Therefore, we have defined intramolecular fragments for all molecules except 1,4-diiodobenzene. The fragment definitions are visualized in Fig. \ref{fig2}.
By dividing molecules into fragments, the SLERP algorithm is applied to each one of them individually and the bond lengths between the fragments are interpolated linearly, ensuring accurate representation of molecular flexibility as well as changes in the bond order. Furthermore, the overall molecular structure is shifted so that the center of mass is translated linearly within the unit cell across the pathway.

Since these fragment interpolations can lead to atomic collision in the unit cell for some images,  we employ a post-processing step that adjusts the unit cell size to ensure chemically reasonable structures throughout the pathway. Therefore, we first detect possible atomic collisions by searching across all images for intermolecular distances smaller than $1.4$ times the respective covalent radius as used by ASE. In the image with the maximum number of collisions we obtain the smallest occurring intermolecular distance. Then, the unit cell vectors for that particular image are increased by the respective projection of the identified distance vector. Next, the linear interpolation of the unit cell vectors is repeated for the whole pathway utilizing the newly changed cell parameters as fixed intermediate image. This process is repeated until the resulting pathway is completely free of collisions. 

This initial interpolation methodology was utilized throughout. 
For 1,4-diiodobenzene we have additionally used a simple linear interpolation for comparison, for the other systems this leads to atomic collisions.

Next, we have performed nudged elastic band (NEB) calculations  utilizing the solid-state NEB method\cite{Sheppard2012} as implemented in TSASE, which allows for variable unit cells during the reaction path sampling. 
We always applied the climbing image method\cite{Henkelman2000} and used as convergence criteria 0.005 eV\,\AA{}$^{-1}$ for 1,4-diiodobenzene and 0.1~eV\,\AA{}$^{-1}$ for the two remaining, more complex systems.

For 1,4-diiodobenzene we have performed NEB calculations with PBE+MBD utilizing the same settings as for the lattice relaxations described above. In this case, we have also performed a harmonic phonon calculation for the obtained transition-state estimate to validate the NEB results, as well as for the optimized structures of both polymorphs. For this calculation, we applied PBE+MBD utilizing phonopy\cite{Togo2015} together with FHI-aims using a supercell of at least 12 \AA{} in each direction to avoid artifacts and  finite displacements of 0.005 \AA{}. The q-grid for the evaluation of vibrational free energies was chosen so that the number of q-points ($n$) in every direction satisfies $n \times l \geq 50$ \AA{}, with $l$ being the unit-cell length in the respective direction. 
Furthermore, we have also performed a NEB pathway optimization for 1,4-diiodobenzene using DFTB3+MBD as implemented in DFTB+\cite{Hourahine2020}, with a k-point grid satisfying $n \times l \geq 18$ \AA{} and utilizing the 3ob parameter set\cite{Gaus2012,Gaus2014,Kubillus2014}.

Next, we have created PBE+MBD reference data for the training of machine-learned force fields (MLFFs) suitable for the NEB calculations of interest. Therefore, we have performed for all systems several short 5 ps molecular dynamics (MD) simulations with the isothermal-isobaric ensemble (NPT) at 500 K utilizing FHI-aims together with ASE. 
We have used a timestep of 1 fs and characteristic timescales of 15 fs and 45 fs for the thermostat and the barostat constant, respectively. For every system, we have performed  separate 5 ps simulations for the initial and final structures as well as for a structure directly in between according to our introduced interpolation. Furthermore, all these simulations were not only performed at ambient pressure but also at a pressure of 1 GPa. We note that the goal of these MD simulations is not to obtain proper dynamics of the systems but rather sample a diverse set of structures close to the ones observed along the respective transition paths we aim to study.
For 1,4-diiodobenzene and theophylline we have utilized a 1x2x2 and 1x3x1 supercell, respectively.

In addition, we have performed full lattice relaxations of all initial reaction-path images directly after interpolation. In this way, the collective reference energies and forces from several MD simulations and optimization trajectories should cover a sufficiently large portion of the needed potential energy surface, suitable for the specific NEB calculations. In order to remove duplicate structures and high-energy structures we have ensured that the final reference data only include structures that have a RMSD of at least 0.04 \AA{} w.r.t. all other structures and from the individual MD trajectories we have removed all structures, which were by more than 1.5 standard deviations less stable than the average energy of the respective trajectory. 

The obtained PBE+MBD data was now used to train for each system an individual SO3krates\cite{frank2022so3krates,frank2024euclidean} MLFF model with a local neighborhood cutoff of 5 \AA{}, applied minimal image convention, a feature dimension of 132, 3 layers, 4 heads in the message-passing update, and degrees $l = \{ 1, 2, 3\}$ in the Euclidean variables for 2000 epochs.
For all three models, we utilized 5000 samples for training and 1000 samples for validation.
Subsequently, lattice relaxations and subsequent NEB calculations were carried out for all systems in the same fashion as described for the PBE+MBD calculations above. We have utilized the GLP\cite{Langer2023} package to obtain ASE calculators of our three MLFFs. 

Finally, we have performed several single-point calculations on top of the obtained optimized NEB reaction paths, starting with PBE+MBD/light single points within FHI-aims (as described above) using the SO3krates structures. For the PBE+MBD/light reaction path of 1,4-diiodobenzene, we have additionally calculated PBE+MBD energies utilizing tight species default settings and ME3(PBE0+MBD:PBE+MBD)/tight multimer embedding\cite{Hoja2023} energies, essentially approximating tight PBE0+MBD results.

\section{Results and Discussion}

\subsection{1,4-Diiodobenzene Polymorph Transition}

We start by discussing the smallest considered system --- 1,4-diiodobenzene. 
The low-temperature $\alpha$ polymorph and the high-temperature $\beta$ polymorph differ mainly in the relative orientation of two of the four molecules within the unit cell. 
The remaining two molecules maintain a similar position and orientation in both crystalline structures. 
Experimentally, the transition temperature from the $\alpha$ to the $\beta$ polymorph amounts to 326 K~\cite{Alcob1994}, suggesting a quite small energy barrier for this transition.

\begin{figure}[!]
\includegraphics[width=\columnwidth]{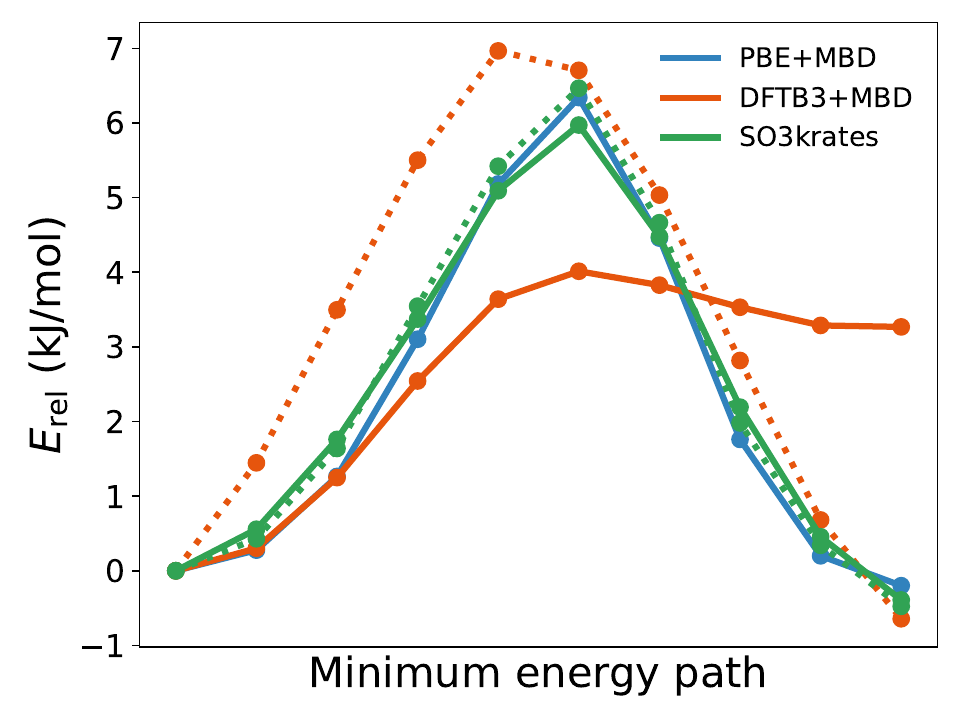}
\caption{\label{fig3} NEB results for the minimum-energy path from polymorph $\alpha$ of 1,4-diiodobenzene (left) to polymorph $\beta$ (right). The solid lines indicate full NEB optimization at that level, while dashed lines indicate PBE+MBD/light single point energies calculated on top of DFTB3+MBD or SO3krates structures. }
\end{figure}

Fig. \ref{fig3} shows the obtained NEB results for several methods for the transition of the $\alpha$ polymorph (left) to the $\beta$  polymorph (right). 
The corresponding energy barriers $\Delta E_{\mathrm{max}}$	(calculated from polymorph $\alpha$ to the maximal energy) and the thermodynamic relative stability of the polymorphs $\Delta E_{\alpha \beta}$, where a negative value indicates that polymorph $\beta$ is more stable, are listed in Table \ref{tab1}.
The small unit cell size and the low complexity of this polymorph transition allow us to perform all necessary calculations fully at the DFT level. 
Hence, we utilize 1,4-diiodobenzene to properly evaluate the performance of our SO3krates MLFF for energies and structures before applying the MLFF models to more complex systems.
The PBE+MBD/light results yield a small energy barrier of only 6.3 kJ\,mol$^{-1}$, which fits well with a transition temperature close to room temperature. 
In terms of thermodynamics, both polymorphs have virtually the same stability ($\Delta E_{\alpha \beta} = -0.2$ kJ\,mol$^{-1}$), since we cannot reliably distinguish between such small stability differences with DFT calculations. 
Please note that the values reported here are relative static lattice energies (normalized per molecular unit), which do not include vibrational free energy effects.
When performing tight PBE+MBD single points or tight PBE0+MBD estimates via multimer embedding\cite{Hoja2023} including up to trimer interactions (ME3), the relative energies change less than 1 kJ\,mol$^{-1}$. 

\begin{table}[!]
\setlength{\tabcolsep}{8pt}
\centering
\caption{Energy differences in kJ\,mol$^{-1}$ for the polymorph transition of 1,4-diiodobenzene normalized per molecular unit.}
\label{tab1}
\begin{tabular}{@{}llrr@{}}
\hline\hline
Energy Method	&	Opt. Method	&	$\Delta E_{\mathrm{max}}$	&	$\Delta E_{\alpha \beta}$\\
\hline
PBE+MBD/light	&	PBE+MBD/light	&	6.3	&	-0.2	\\
PBE+MBD/tight	&	PBE+MBD/light	&	5.4	&	-0.4	\\
ME3/tight$^a$ 	&	PBE+MBD/light	&	6.4	&	 0.0	\\
DFTB3+MBD	    &	DFTB3+MBD	    &	4.0	&	 3.3	\\
PBE+MBD/light	&	DFTB3+MBD	    &	7.0	&	-0.6	\\
SO3krates	    &	SO3krates	    &	6.0	&	-0.4	\\
PBE+MBD/light	&	SO3krates	    &	6.5	&	-0.5	\\	
\hline\hline
\end{tabular}\\
$^a$ ME3(PBE0+MBD:PBE+MBD)/tight 
\end{table}

When analyzing the NEB results obtained with the semi-empirical density functional tight binding approach (DFTB3+MBD), one can clearly see that there are significant differences compared to PBE+MBD.
The energy barrier is with 4 kJ\,mol$^{-1}$ much smaller and the energy difference between the two polymorphs is now much larger, amounting to more than 3 kJ\,mol$^{-1}$. 
When we utilize the DFTB3+MBD structures and perform on top of them PBE+MBD/light single-point calculations, we get similar results to the PBE+MBD NEB results, but the maximum is slightly shifted. 
This suggests that DFTB3+MBD is not reliable in this context when it comes to energies, but the structures along the minimum energy path seem to have a better quality.

Next, we evaluate the performance of our SO3krates MLFF.
When we compare the $\Delta E_{\mathrm{max}}$ and $\Delta E_{\alpha \beta}$ values obtained with SO3krates with those of the fully optimized PBE+MBD/light calculations, both values differ by less than 0.3 kJ\,mol$^{-1}$.
Also, the maximum along the minimum-energy path is located at the same image in both cases and the energy difference of the two methods for the individual points along the path remains consistently below 0.6 kJ\,mol$^{-1}$.
Hence, SO3krates is able to very accurately reproduce the PBE+MBD energies of this transition. 
Additionally, we have performed PBE+MBD single point calculations on top of the SO3krates structures to evaluate the accuracy of the energies without the influence of geometry effects.
In this case, all differences between the SO3krates results and the PBE+MBD single points (on the same structures) along the minimum energy path are less than 0.5~kJ\,mol$^{-1}$ and we obtain a mean absolute error (MAE) and a mean error (ME) for the relative energies of only 0.2 and -0.1 kJ\,mol$^{-1}$, respectively. 
Performing the NEB calculations with 15 instead of the otherwise used 10 images yields virtually identical results (within 0.1 kJ\,mol$^{-1}$).
Hence, our SO3krates MLFF yields very accurate results and significantly outperforms DFTB3+MBD. Therefore, we have performed the full NEB calculations for the remaining two systems only with the generated MLFF.

However, the so far discussed accuracies of the obtained SO3krates MLFF were only based on a few energy differences. 
Therefore, Table \ref{tab2} lists the accuracy of the MLFF for a large test set containing 7125 1,4-diiodobenzene structures, which were not used for training or validation.
It can be seen that the obtained energies and forces are highly accurate.
The MAE of the energies amounts to only 0.13 kJ\,mol$^{-1}$ per molecular unit or 0.01 kJ\,mol$^{-1}$ when expressed per atom, and the MAE of the forces amounts to only 1.1 kJ\,mol$^{-1}$\,\AA{}$^{-1}$, which corresponds to a coefficient of determination of 0.9998. To put the force error into perspective, it corresponds to 0.011 eV\,\AA{}$^{-1}$, which is only about twice as large as our (quite strict) convergence setting used for the DFT optimizations (0.005 eV\,\AA{}$^{-1}$). 
 
\begin{table}[!]
\setlength{\tabcolsep}{6pt}

\centering
\caption{Mean absolute errors (MAE) and coefficients of determination (R$^2$) for SO3krates energies and atomic forces for the respective test sets compared to PBE+MBD/light reference data. Energies are given in kJ\,mol$^{-1}$ (normalized per molecular unit) and forces are in kJ\,mol$^{-1}$\,\AA{}$^{-1}$. For the Diels-Alder reaction the molecular unit refers to one bis(N-allylimino)-1,4-dithiin/9-bromoanthracene co-crystal unit or the resulting Diels-Alder product. }
\label{tab2}
\begin{tabular}{@{\extracolsep{6pt}}lrrrr@{}}
\hline\hline
 & \multicolumn{2}{c}{Energies} & \multicolumn{2}{c}{Forces}\\
 \cline{2-3} \cline{4-5}
 System & MAE & R$^2$ & MAE  & R$^2$\\
\hline
1,4-Diiodobenzene       & 0.13  & 0.9998 & 1.10 & 0.9998\\
Theophylline            & 0.26  & 0.9999 & 1.68 & 0.9994\\
Diels-Alder Reaction    & 0.75  & 0.9996 & 1.92 & 0.9993\\
\hline\hline
\end{tabular}\\
\end{table}

Since it is possible for this system to utilize a simple linear interpolation as starting point of NEB, we have compared the efficiency with our new interpolation method. 
Both methods yield identical results, but our new methodology converged slightly faster, needing 12~\% fewer steps than the simple linear interpolation. 

After having discussed the energetics, we analyze the obtained transition-state geometries. 
The unit cell of the PBE+MBD transition state is visualized in Fig. \ref{fig1}. 
It can be seen that the orientation of the left two molecules is exactly in between both polymorphs. 
To assess the quality of our PBE+MBD transition-state geometry, we have calculated the phonon frequencies and find two imaginary frequencies at the $\Gamma$-point: -73.6 cm$^{-1}$, which corresponds to the movement along the NEB pathway, and another mode at -14.6 cm$^{-1}$. 
At a true transition state, we should only observe the first imaginary mode. 
Therefore, we have slightly displaced the structure along the second imaginary mode, followed by a subsequent phonon calculation.
In this way, we were able to fully remove this second imaginary mode and have obtained a true transition state for this polymorph transition.
The differences between the two structures are essentially negligible, with a lattice energy difference of 0.01 kJ\,mol$^{-1}$ and a RMSD$_{20}$ value of 0.038 \AA{}, which is the root mean-square deviation (RMSD) of a cluster consisting of 20 molecules.

In order to illustrate the effect of vibrational free energies on $\Delta E_{\mathrm{max}}$ and $\Delta E_{\alpha \beta}$, we have additionally performed PBE+MBD/light phonon calculations for the optimized structures of polymorphs $\alpha$ and $\beta$. 
At room temperature, the resulting free energy difference in polymorph stability amounts to -0.21 kJ\,mol$^{-1}$. 
This is only a very small shift by 0.01 kJ\,mol$^{-1}$ compared to the static lattice energies, but we note that for other systems the inclusion of vibrational free energies can often be crucial for determining the correct thermodynamic stabilities\cite{Hoja2019}.
The calculated free energy barrier  amounts at room temperature to 5.4 kJ\,mol$^{-1}$, which is about 1 kJ\,mol$^{-1}$ lower than the corresponding static lattice energy result.

When we compare the obtained transition states of DFTB3+MBD and SO3krates with the PBE+MBD results, we obtain RMSD$_{20}$ values of 0.308 \AA{} and 0.199~\AA{}, respectively. 
Also in this case, the SO3krates MLFF outperforms DFTB3+MBD and yields a structure very close to the PBE+MBD result. 
A detailed analysis of the accuracy of the MLFF for phonon calculations and transition state optimizations will follow in a subsequent publication.

 \subsection{Theophylline Polymorph Transition}

Next, we discuss the polymorphic transition from theophylline  polymorph I~\cite{Khamar2011}, which is stable at higher temperatures,  to the room-temperature stable polymorph II~\cite{https://doi.org/10.5517/ccygrw0}. 
This transition (I to II) does not occur upon cooling due to significant structural differences.\cite{Fucke2012} 
However, the reverse transition of polymorph II into I is known to occur upon heating.
Since we can expect here a lot of atomic motion, this system is ideal for testing our new interpolation method. 
When dividing every molecule into three fragments (see Fig. \ref{fig2}), our new interpolation method automatically creates initial reaction path structures without any atomic collisions, while this is not possible with a standard linear interpolation. 

\begin{figure}[!]
\includegraphics[width=\columnwidth]{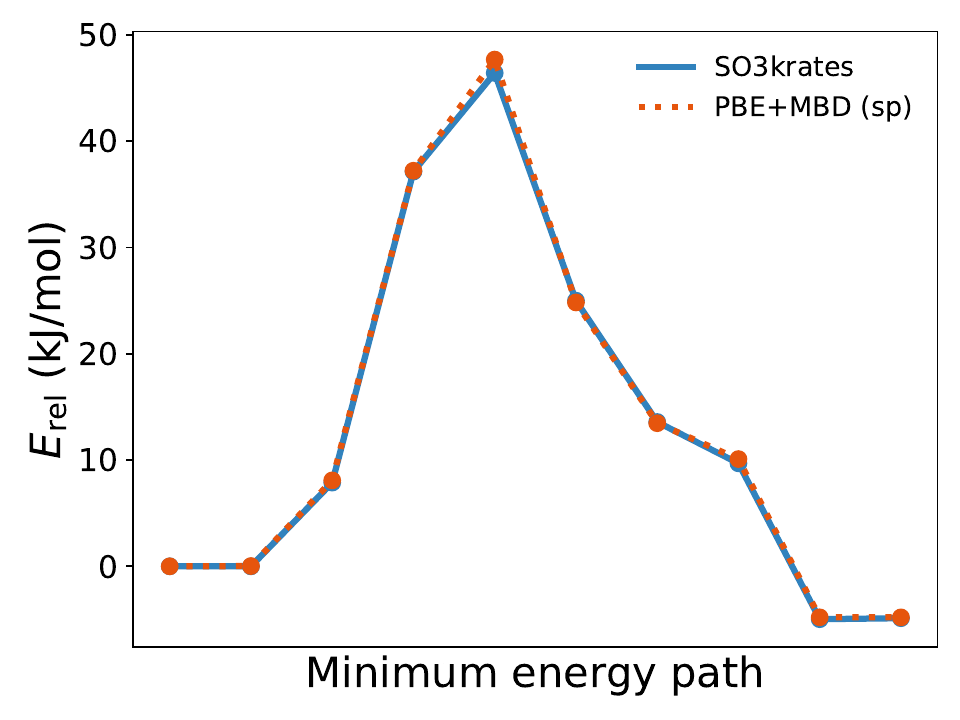}
\caption{\label{fig4} NEB results for the minimum-energy path from polymorph I of theophylline  (left) to polymorph II (right). The solid lines indicate full NEB optimization, while the dashed lines indicate PBE+MBD/light single point energies calculated on top of the SO3krates structures. }
\end{figure}

The results of the minimum-energy path obtained with the SO3krates  MLFF together with single point calculations on top of the SO3krates structures are shown in Fig. \ref{fig4}. 
In both cases, polymorph II is, as expected, thermodynamically more stable by about 5 kJ\,mol$^{-1}$ and we obtain a substantial energy barrier, amounting to 46.4 kJ\,mol$^{-1}$ and 47.7 kJ\,mol$^{-1}$ for SO3krates and the PBE+MBD single points, respectively. 
The SO3krates MLFF provides again very accurate results, underestimating the PBE+MBD energy barrier by only 3 \% and  yielding MAE and ME values across the minimum energy path of 0.3 and -0.2 kJ\,mol$^{-1}$, respectively.
The obtained transition state is shown in Fig. \ref{fig1} and one can see that this process involves substantial motion of all molecules, given the large changes in unit-cell parameters.

The MAE of our theophylline SO3krates MLFF across a test set containing 9276 structures amounts for energies to 0.26 kJ\,mol$^{-1}$ when normalized per molecule (see Table \ref{tab2}) and again 0.01 kJ\,mol$^{-1}$ when normalized per atom. 
The MAE of the forces is with 1.68 kJ\,mol$^{-1}$\,\AA{}$^{-1}$ only slightly higher than that for 1,4-diiodobenzene.

\subsection{Solid-State Diels-Alder Reaction}

\begin{figure}[!]
\includegraphics[width=\columnwidth]{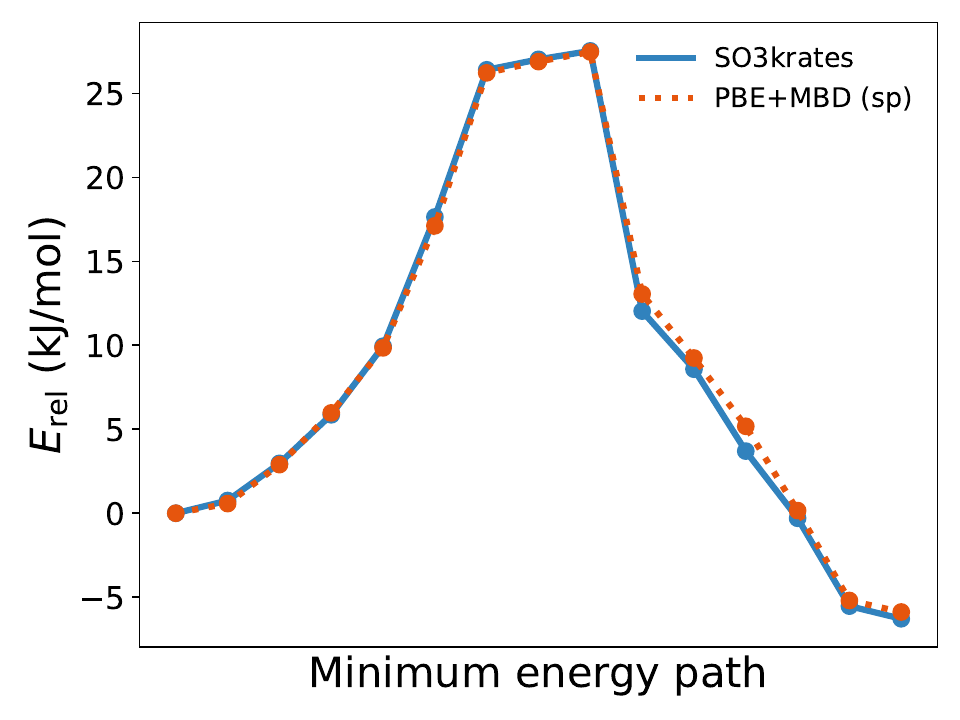}
\caption{\label{fig5}NEB results for the minimum-energy path from a co-crystal of bis(N-allylimino)-1,4-dithiin and 9-bromoanthracene (left) to the respective Diels-Alder product (right). The solid lines indicate full NEB optimization, while the dashed lines indicate PBE+MBD/light single point energies calculated on top of the SO3krates structures.}
\end{figure}

Finally, we investigate if our approach can also be applied to solid-state reactions. 
As an example we utilize a single-crystal to single-crystal Diels-Alder reaction  of a co-crystal composed of bis(N-allylimino)-1,4-dithiin and 9-bromoanthracene.\cite{Khorasani2015} 
Experimentally, it took 84 days to achieve a 97 \% conversion. 

When dividing both molecules building up the co-crystal into several fragments (see Fig. \ref{fig2}), our interpolation yielded also for this reaction automatically collision-free initial path structures.
Fig. \ref{fig5} shows the results of the minimum-energy path obtained with the SO3krates MLFF together with subsequent single point calculations on top of the SO3krates structures. 
In both cases, the product of the reaction is about 6.0 kJ\,mol$^{-1}$ more stable than the co-crystal formed by the reactants. 
The energy barrier is for SO3krates and PBE+MBD virtually identical and amounts to 27.5 kJ\,mol$^{-1}$ in both cases.
When comparing the differences between the two methods for the individual images along the obtained minimum energy path, we find a maximum deviation of 1.5~kJ\,mol$^{-1}$ and the MAE and ME amount to only 0.4 and  -0.2~kJ\,mol$^{-1}$, respectively.
Hence, our obtained SO3krates MLFF is also able to describe the energetics along the minimum energy path of a solid-state reaction with high accuracy. 

The corresponding transition state is visualized in Fig.~\ref{fig1}.
Before the actual ring closure can happen, several propenyl groups need to rearrange themselves, which apparently is the main source of the energy barrier in this case.

After having discussed the specific minimum energy path for the reaction, we illustrate the accuracy of the SO3krates model for a large test set containing 8165 structures (see Table \ref{tab2}). 
This test set includes co-crystal structures of bis(N-allylimino)-1,4-dithiin/9-bromoanthracene as well as structures containing the respective Diels-Alder product. 
Therefore, the molecular unit we have used to normalize the energies corresponds to one product molecule or one co-crystal unit.
The resulting MAE amounts to 0.75 kJ\,mol$^{-1}$, which is very accurate, especially given the already quite complicated system. 
While the error appears larger than that for the other two systems in this notation (normalized per molecular unit), it is virtually identical to that of the other systems when normalized per atom (0.01 kJ\,mol$^{-1}$). 
The forces have with 1.92 kJ\,mol$^{-1}$\,\AA{}$^{-1}$ a slightly larger MAE than the two previous systems, but are with a coefficient of determination of 0.9993 still very accurate.

\section{Conclusion}
We have introduced an interpolation method for generating an initial pathway for subsequent nudged-elastic band (NEB) calculations suitable for periodic structures like molecular crystals. 
This methodology aims to automatically avoid any atomic collisions, which otherwise regularly happen in periodic systems when using simple linear interpolations. 
We demonstrated the versatility of this approach by successfully applying it to a range of diverse systems, including two polymorph transitions --- one with a small and one with a rather large energy barrier --- as well as a single-crystal Diels-Alder reaction. 
Our approach consistently provided collision-free initial pathways, avoiding the need for cumbersome manual corrections. 
This automated pathway generation greatly simplifies the initial setup for NEB calculations, making it a highly practical tool for tackling even complex periodic systems.

Furthermore, we have utilized the SO3krates architecture to train an individual machine-learned force field (MLFF) for every studied system.
By training the MLFFs on a diverse set of reference data obtained from PBE+MBD calculations for the specific systems, we are able to reproduce the energy barriers and structural characteristics of the minimum energy pathways with high accuracy. 
The MAEs between the MLFF energies along the reaction paths compared to subsequent PBE+MBD single-point energies are for all three systems smaller than 0.4 kJ\,mol$^{-1}$ and the maximal error for individual images amounts to less than 1.5 kJ\,mol$^{-1}$.
In terms of our respective test sets containing in all cases more than 7000 structures calculated with PBE+MBD, all our MAEs for energies normalized per molecular unit amount to less than 0.8 kJ\,mol$^{-1}$ or when expressed as MAE per atom consistently yield an error of 0.01 kJ\,mol$^{-1}$. 
In terms of the forces, all three MAEs are below 2 kJ\,mol$^{-1}$\,\AA{}$^{-1}$.
However, we note that the reference data focused heavily on the PES region required for the studied transition and the resulting MLFFs will likely be less accurate when simulating for instance other polymorphs.

Finally, many polymorph transition and reaction paths can probably not be described by utilizing unit cells but would require larger supercells to allow for the needed flexibility. 
In such cases, computationally efficient methods like MLFFs are key for being able to afford the necessary calculations. 
Our results highlight that with properly trained MLFFs, it is possible to achieve both computational efficiency and accuracy, making them a very promising tool for exploring complex solid-state phenomena. 
Therefore, the next steps will be to apply this methodology to NEB calculations involving larger supercells, to utilize these MLFFs for explicit molecular dynamics simulations, and to test their applicability for vibrational free energy calculations of molecular crystals.

\begin{acknowledgments}
The computational results presented have been achieved in part using the Vienna Scientific Cluster (VSC) and we thank Prof. A. Daniel Boese for providing resources at the University of Graz.
\end{acknowledgments}

\end{document}